\documentclass[twocolumn,aps,superscriptaddress,prd,10pt]{revtex4-1}
\usepackage{multirow}
\usepackage{graphicx}
\usepackage{dcolumn}
\usepackage{bm}
\usepackage{bbold}
\usepackage{braket}
\usepackage{amssymb,amsmath}
\usepackage[section]{placeins}
\usepackage{slashed}

\usepackage{color}
\usepackage{amsfonts}
\usepackage{subfigure}
\usepackage{array}
\usepackage{enumerate}
\usepackage{cancel,soul,ulem}
\renewcommand{\sout}{\bgroup \color{red} \ULdepth=-.5ex \ULset}
\usepackage{xspace}
\usepackage{siunitx}
\usepackage{xfrac}
\usepackage{hyperref}
\usepackage[nameinlink]{cleveref}
\usepackage{appendix}
\usepackage{subfigure}
\usepackage{xifthen}
\usepackage{xcolor}
\hypersetup{
	colorlinks,
	linkcolor={red!75!black},
	citecolor={blue!75!black},
	urlcolor={blue!75!black}
}
\graphicspath{{./figures/}{./}}
\begin{document}
\title{False-Vacuum-Trapped Neutron-Rich Droplets from the Nuclear Liquid-Gas Transition}
\author{Jingdong Shao} 
\email[]{shaojingdong19@mails.ucas.ac.cn} 
\affiliation{School of Physical Sciences, University of Chinese Academy of Sciences, Beijing 101408, China} 
\author{Mei Huang}
\email[]{huangmei@ucas.ac.cn (corresponding author)}
\affiliation{School of Nuclear Science and Technology, University of Chinese Academy of Sciences, Beijing 101408, China}

\begin{abstract}
Subsaturation liquid-gas transitions in heavy-ion collisions undergo spinodal decomposition, seemingly precluding metastable false vacuum relics. We propose that novel false-vacuum-trapped neutron-rich droplet relics emerge via cluster quenching therein, populating uncharted nuclear regions. Negative pressure confines neutrons beyond the drip line with energetically forbidden single-neutron emissions. Isospin barriers sustain nucleation lifetimes over microseconds for an appreciable part of droplets. Accessible yields and low charge-to-mass ratios enable detection via fragment spectroscopy.
\end{abstract}  

\maketitle
{\it Introduction.}
Trapping matter in a metastable false vacuum produces long-lived states across condensed matter, nuclear physics, and cosmology \cite{angell1982supercooled,debenedetti2001supercooled,angus1988low,bundy1989pressure,levi1978dynamics,dracoulis2016review}. In particular, abnormal nuclear states associated with nontrivial false vacua have been proposed \cite{kolomeitsev2026abnormal}, including Lee and Wick's chiral-restored matter at high baryon density $n\gg n_0$ (where $n_0$ is the nuclear saturation density) \cite{Lee:1974ma}, Migdal's pion-condensed nuclei \cite{RevModPhys.50.107,MIGDAL1990179,MIGDAL1976423,MIGDAL1974172}, Witten's strange quark matter \cite{Witten:1984rs}, and Bodmer's collapsed nuclei \cite{PhysRevD.4.1601}. Yet searches at heavy-ion collision (HIC) facilities have found no conclusive evidence for such extreme-density states. At lower densities, experimental signatures such as bimodal distributions and Fisher scaling \cite{bonnet2009bimodality,moretto1997fisher} strongly support a first-order nuclear liquid-gas phase transition (LGPT) \cite{Borderie2019}. During radial expansion, however, the bulk fireball rapidly enters the spinodal instability region, which amplifies density fluctuations \cite{steinheimer2012spinodal} and drives spinodal multifragmentation \cite{chomaz2004nuclear,borderie2008nuclear,borderie2001evidence,Matera:2003ap}.

The homogeneous scenario, however, omits the strong spatial fluctuations and dynamical nucleon-nucleon correlations inherent in the microscopic evolution \cite{Aichelin1991}. Beyond the mean-field approximation \cite{Colonna2020}, highly correlated local clusters \cite{Ono2019,Ono2004} can dynamically decouple from the collective flow during the early expansion stage, thereby avoiding rapid dilution and spinodal instability. Measurements with $4\pi$ detector arrays corroborate this survival mechanism: mid-rapidity intermediate-mass fragment (IMF) enhancement, kinematic decoupling, and anisotropic angular distributions \cite{10.3389/fphy.2022.1050450,de2012correlations,baran2004neck} show that early-emitted dynamical clusters deviate from statistical fragmentation \cite{DeFilippo2014}. The decoupled clusters also exhibit neutron enrichment \cite{li1998isospin} generated via neutron migration or fractionation \cite{fable2023experimental,baran2004neck,di2009isospin,chen2003light}. 

Based on these features, we propose that thermal fluctuations can excite such neutron-rich clusters into the liquid phase and rapid cooling of the ambient fireball can dynamically quench and trap them in the metastable false vacuum of the LGPT. Ambient neutron enrichment supplies the neutrons required to form these clusters, which subsequently relax into self-bound droplets stabilized by the equilibrium among the nuclear pressure difference, Coulomb repulsion, and surface tension, while coalescence may further enlarge their size. Simultaneously, the isospin asymmetry induces a potential barrier that suppresses decay channels and yields long lifetimes. These false-vacuum-trapped neutron-rich droplets (FNDs) carry characteristic charge-to-mass ratios that can be directly probed at the relevant energy scales of $\mathrm{Au}+\mathrm{Au}$ \cite{PhysRevLett.75.1040,FOPI:1996cjz} and $\mathrm{Sn}+\mathrm{Sn}$ \cite{lee2022isoscaling} at the Heavy Ion Research Facility in Lanzhou (HIRFL) \cite{xu2010nuclear}, the High Intensity Heavy-ion Accelerator Facility (HIAF) \cite{zhou2022status}, Facility for Antiproton and Ion Research (FAIR) \cite{fortov2012european}, Facility for Rare Isotope Beams (FRIB) \cite{balantekin2014nuclear}, Grand Acc\'{e}l\'{e}rateur National d'Ions Lourds (GANIL) \cite{farget2025applications}, and RIKEN Radioactive Isotope Beam Factory (RIBF) \cite{okuno2020present}.

In this Letter, we first establish the LGPT phase diagram and the false vacuum window. After introducing the quenching, we determine the mechanical equilibrium and bulk properties of FNDs, and discuss their lifetimes and experimental signatures in HICs. Finally we extend this scenario to compact stars.

{\it The phase diagram.}
The first-order LGPT is governed by the grand potential density $\Omega$, or equivalently the pressure $P = -\Omega$. The global minimum of $\Omega$ defines the stable true vacuum, whereas a local minimum separated by a potential barrier defines a metastable false vacuum. Although a static first-order transition occurs at the transition temperature $T_c$ where the two vacua are pressure-degenerate, a dynamical transition does not occur immediately at $T_c$ because the barrier suppresses the transition rate. By continuity, there exists a window near $T_c$ where the free-energy difference between the two vacua $\Delta\Omega$ remains so small that the transition rate stays suppressed and that the system requires a long time to tunnel through the barrier. Within this window, the system can be trapped in a metastable false vacuum at a pressure lower than that of the true vacuum. Long-lived false-vacuum trapping is therefore a natural, model-independent feature of a first-order phase transition.

Specifically, at zero temperature the gas phase forms a dilute nuclear gas with $P_\mathrm{g}\sim0$, whereas the liquid phase satisfies $P_\mathrm{l}\sim0$ only at the saturation density $n_0=0.16\text{ fm}^{-3}$. The false-vacuum window therefore appears at subsaturation densities $n \lesssim n_0$, where the false-vacuum pressure is negative, $P_\mathrm{l} < P_\mathrm{g} \sim 0$. Throughout this Letter, "pressure" refers exclusively to the strong-interaction component.

\begin{figure}
  \centering
  \includegraphics[width=0.97\linewidth]{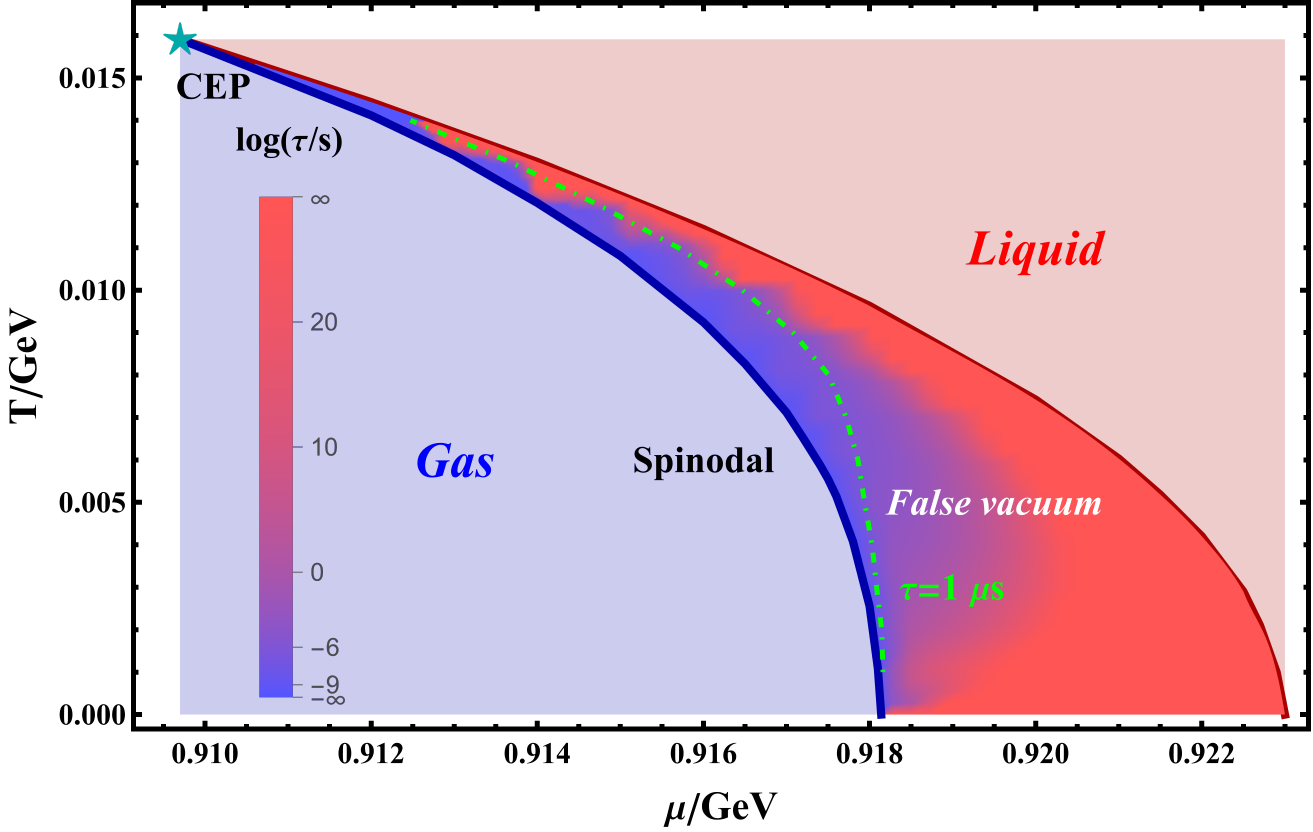}
\caption{The phase diagram of the isospin-symmetric LGPT in the temperature--baryon chemical potential ($T$--$\mu$) plane. The red solid curve bounding the liquid phase denotes the static first-order phase transition line, terminating at the critical endpoint (cyan star). The blue solid curve bounding the gas phase denotes the liquid spinodal line. The false vacuum region is color-coded by the logarithmic lifetime $\log(\tau/s)$ for a sphere with a radius of $10$ fm. The green dashed curve indicates the $\tau = 1~\mathrm{\mu s}$ lifetime contour.}
  \label{phase}
\end{figure}
To quantify this model-independent scenario, we compute the LGPT phase diagram with the two-flavor isospin-symmetric nonlinear Walecka model (see Supplemental Material (SM) for details) in Fig.~\ref{phase}. The red solid boundary of the liquid phase marks the static first-order phase transition line, which terminates at the critical endpoint (CEP) at $\mu = 0.9097$~GeV and $T = 0.01591$~GeV (cyan star). The blue solid boundary of the gas phase denotes the liquid spinodal line, below which the potential barrier vanishes. Between these two boundaries lies the metastable false vacuum window. The color gradient encodes the lifetimes $\log(\tau/s)$ for a sphere with a radius of $10$ fm, and indicates that the false vacuum lifetime increases from $0$ at the spinodal boundary to infinity at the static phase transition line. Particularly, the wide region between the green dashed contour ($\tau = 1~\mathrm{\mu s}$) and the first-order phase transition line exhibits lifetimes exceeding $1~\mathrm{\mu s}$, indicating a broad thermodynamic window supporting a long-lived false vacuum.

The nontrivial false vacuum corresponds to the liquid phase, and the phase diagrams reveal metastable windows at both vanishing isospin chemical potential $\mu_I=0$ in Fig.~\ref{phase} and at finite $\mu_I$ (see Fig. 1 in SM), within which a substantial part of the false vacuum is long-lived. Abnormal nuclear matter therefore need not be restricted to stable states at extreme densities or to isospin-symmetric systems; it can exist within the long-lived false vacuum window at subsaturation.

{\it Droplet.} In HICs, the false vacuum manifests as finite droplets rather than infinite bulk matter. While the expanding homogeneous fireball undergoes spinodal decomposition, strongly correlated neutron-rich clusters form during the early expansion stage and dynamically decouple from the bulk flow. Intense local thermal fluctuations can excite these clusters across the potential barrier into the liquid phase. As the fireball expands and dilutes, the local temperature drops rapidly, depriving the clusters of the thermal energy required to cross the barrier back to the true vacuum and trapping them in the false vacuum. Subsequently they may be enlarged by coalescence. In neutron-rich environments generated by neutron migrations or fractionation, the clusters are protected against the LGPT by the potential barrier induced by $\mu_I$. Since the temperature relevant to HICs $T\sim5\,\,\mathrm{MeV}\ll\mu$ is not important at high density and decreases rapidly during quenching, we focus primarily on products at $T=0$.

\begin{figure}
  \centering
  \includegraphics[width=0.97\linewidth]{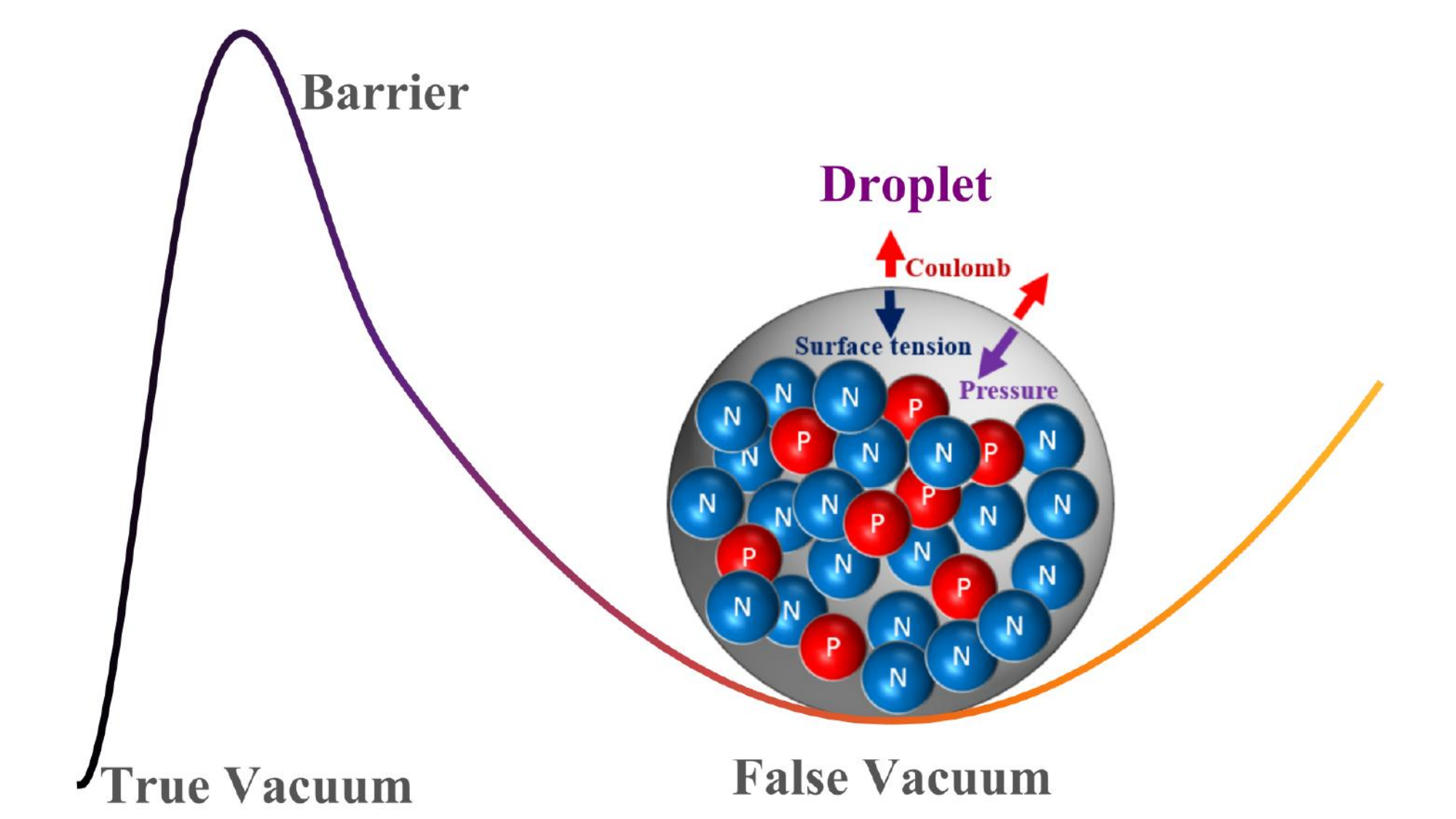}
  \caption{The sketch of FNDs (not to scale).}
  \label{sketch}
\end{figure}

Because the trapped false vacuum has lower pressure than the true vacuum gas, the inward pressure difference together with short-range cohesive surface tension counterbalances long-range Coulomb repulsion, and the mechanical equilibrium stabilizes the quenched clusters at finite equilibrium radii as FNDs. As illustrated in Fig.~\ref{sketch}, rapid quenching, neutron enrichment and mechanical equilibrium stabilize FNDs in the false vacuum.

To evaluate the FND properties quantitatively, we adopt a liquid-drop model. In a finite droplet, surface and Coulomb terms also contribute to the total grand potential density $\Omega_t$
\begin{equation}
\Omega_t=\Omega+\frac{3\sigma_\mathrm{eff}}{R}+\frac{9Z^2e^2}{80\pi^2R^4},
\end{equation}
where $Z = \frac{4}{3}\pi R^3 n_p$ is the proton number, and $\sigma_\mathrm{eff}$ is the order-parameter- and isospin-dependent \cite{1991NuPhA} effective surface tension (see SM for details). Equilibrium among the internal and external nuclear pressures $P_{\mathrm{in}}$ and $P_{\mathrm{out}}$ against the outward Coulomb pressure and the inward Laplace pressure from surface tension yields \cite{chamel2008physics}
\begin{equation}
  P_{\mathrm{in}}+\frac{1}{15}e^2n^2_pR^2=P_{\mathrm{out}}+\frac{2\sigma_\mathrm{eff}}{R},
\end{equation}
where $R$ is the droplet radius, $n_p$ is the proton number density and $e=0.30286$ is the unit charge. 

The external pressure approximately vanishes $P_{\mathrm{out}}\sim0$ at $T=0$ for products at detectors, while at finite temperature $T\sim5$ MeV relevant to HICs, $P_{\mathrm{out}}$ is provided by the free neutron gas that maintains chemical equilibrium with neutrons in the droplets. The internal pressure  $P_\mathrm{in}=-\Omega$ is obtained in the liquid phase by treating the nucleons as a thermodynamic system. 

In HICs, droplets need not achieve $\beta$-equilibrium before reaching detectors because the timescale is too short for weak interactions. Products with velocities near the speed of light reach detectors over meter-scale distances within a nanosecond-scale time, while the characteristic time of $\beta$-equilibrium is of the millisecond or longer for nuclides or nuclear matter at $T\sim5$ MeV  \cite{PhysRevC.93.025805,PhysRevC.109.055803}. Conservatively, droplets with lifetimes of microseconds scale or longer can therefore reach detectors. Meanwhile, the eventual states of droplets constrained by $\beta$-equilibrium $\mu_n-\mu_p=\frac{2}{5}e^2n_pR^2$ are for reference.
\begin{figure}
  \centering
\includegraphics[width=0.97\linewidth]{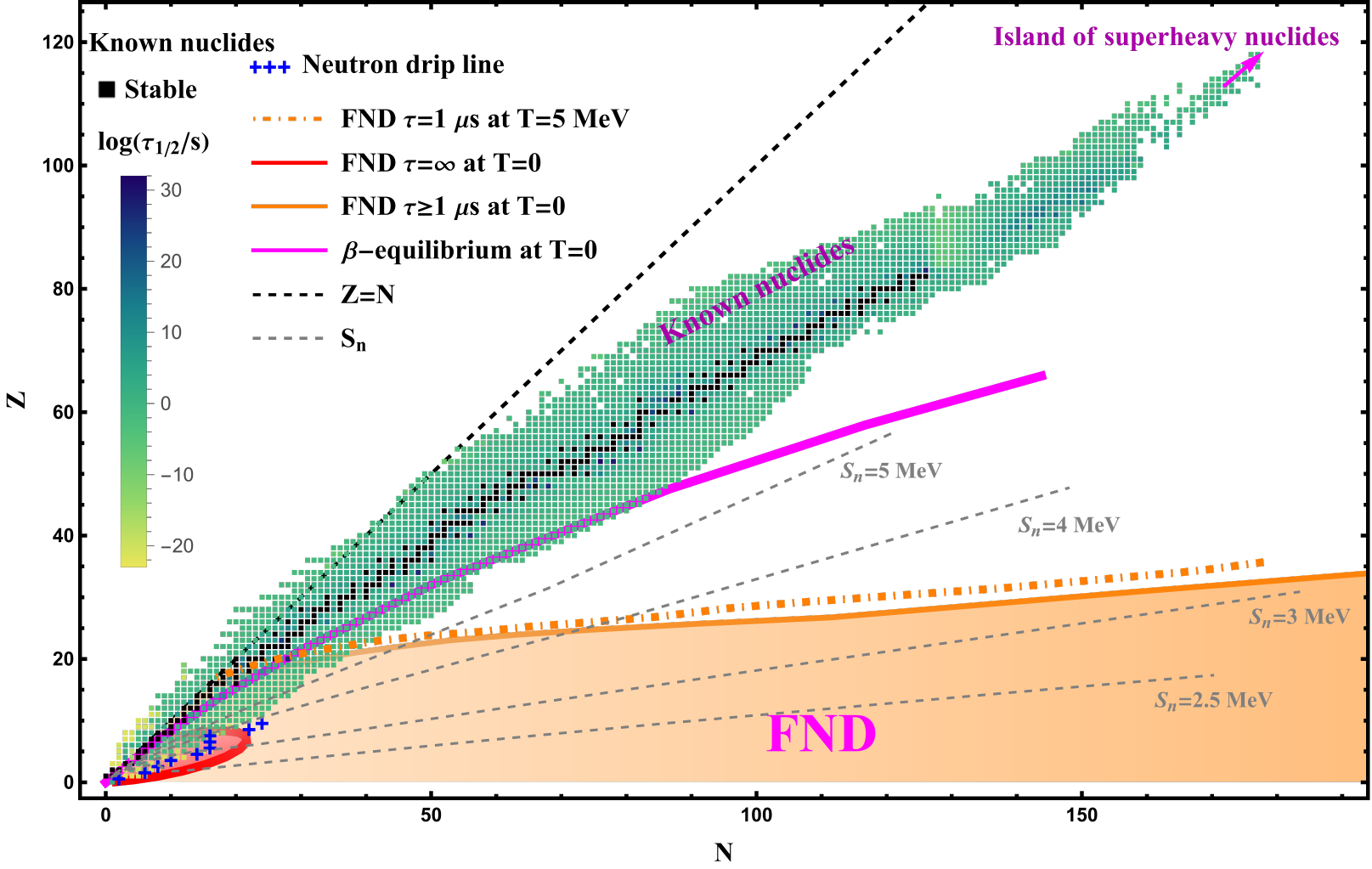}
  \caption{Predicted FND regions compared with known nuclides in the proton--neutron ($Z$--$N$) plane. Known nuclides are color-coded by their logarithmic half-lives $\log(\tau_{1/2}/\mathrm{s})$, with black squares marking stable nuclei, the black dashed line indicating $Z=N$, and the blue crosses denoting the conventional neutron drip line. The orange shaded region bounded by the orange curve marks FNDs with lifetimes $\tau \ge 1~\mathrm{\mu s}$ against false vacuum decay at $T=0$, while the dashed orange curve marks FNDs with $\tau=1~\mathrm{\mu s}$ at $T=5$ MeV. The region bounded by the red curve at low $N$ and $Z$ marks stable FNDs ($\tau = \infty$) at $T=0$. The magenta solid curve shows the FND trajectory satisfying Coulomb-corrected $\beta$-equilibrium, and the gray contours denote different neutron separation energies $S_n$.}
  \label{NZ}
\end{figure}

Figure \ref{NZ} compares the FND regions to the known nuclides \cite{kondev2021nubase2020} with color-coded half-lives $\log(\tau_{1/2}/\mathrm{s})$ and the neutron drip line \cite{NNDC_NuDat3} (blue) in the proton--neutron ($Z$--$N$) plane. FNDs with lifetimes $\tau \ge 1~\mathrm{\mu s}$ extend deeply into the orange shaded neutron-rich region, while the dashed orange curve marks the boundary of FNDs with lifetimes $\tau\ge1~\mathrm{\mu s}$ at $T=5$ MeV. The small gap between the solid and dashed orange curves verifies that temperature affects boundaries weakly. Because the region of FNDs with $\tau\ge1~\mathrm{\mu s}$ at $T=5$ MeV contains the one at $T=0$, FNDs in the orange shaded region must also satisfy $\tau\ge1~\mathrm{\mu s}$ at $T=5$ MeV, and thus the detectable region is not compromised by finite temperature relevant to HICs. Here lifetimes refer only to those of the false vacuum decay in the LGPT, and we will discuss them later. A pocket of stable FNDs with $\tau = \infty$ lies around the conventional drip line. Furthermore, the magenta curve denotes the eventual states of FNDs that satisfy the Coulomb-corrected $\beta$-equilibrium for reference. The gray contours show different neutron separation energies $S_n$ (see SM) above $0$. FNDs cannot be interpreted as an extrapolation of ordinary neutron-rich nuclei; they populate an otherwise empty region and provide a target for identifying them through event-by-event fragment spectroscopy.

\begin{figure}
  \centering
  \includegraphics[width=0.97\linewidth]{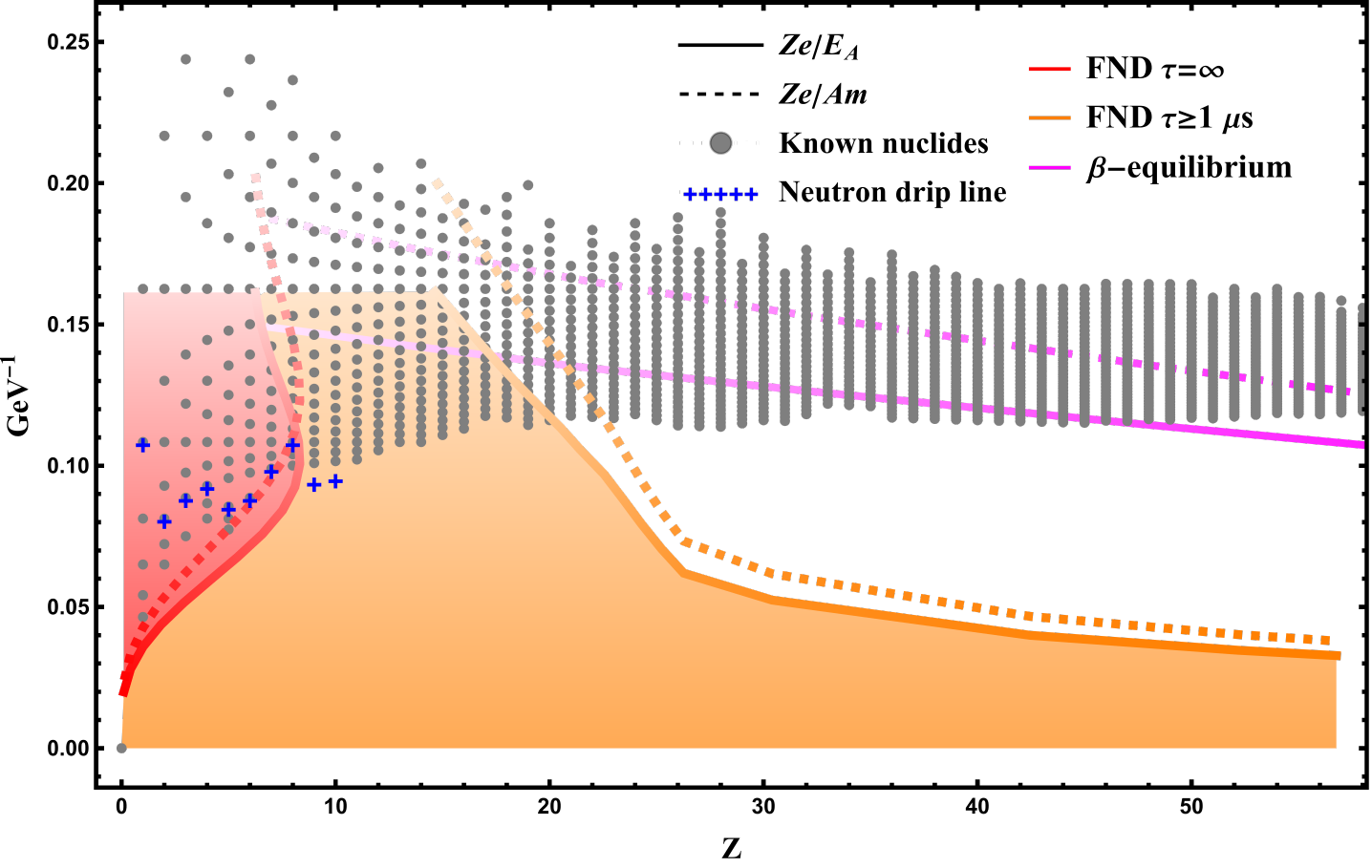}
  \caption{Charge-to-mass ratios of FNDs compared to those of known nuclides (gray) and the neutron drip line (blue) as a function of $Z$. The solid and dashed curves denote the charge-to-energy and charge-to-effective-mass of FNDs, respectively. The orange and red shaded regions denote FNDs with lifetimes $\tau \ge 1~\mathrm{\mu s}$ and $\tau = \infty$, respectively, while the magenta curves denote the trajectories with $\beta$-equilibrium.}
  \label{charge_mass_ratio}
\end{figure}

Figure \ref{charge_mass_ratio} compares charge-to-energy $\frac{Ze}{E_A}$ (solid) and charge-to-effective-mass $\frac{Ze}{Am}$ (dashed) ratios of FNDs to the charge-to-mass ratios of known nuclides (gray) and the neutron drip line (blue). The separation between the solid and dashed curves indicates that the total energy receives comparable contributions from both effective masses and false-vacuum energy. In the stable region (red shaded), ratios of FNDs coincide with those around the neutron drip line. In contrast, ratios of FNDs with $\tau \ge 1~\mathrm{\mu s}$ (orange shaded) are anomalously smaller than those of known nuclides owing to the neutron enrichment. The resulting low but nonzero ratios indicate that FNDs would appear as heavy, positively charged objects, and would therefore exhibit high magnetic rigidity $B\rho$ at given velocity and long times of flight. 

The corresponding energy and density ranges can be probed in $\mathrm{Au}+\mathrm{Au}$ and $\mathrm{Sn}+\mathrm{Sn}$ experiments by worldwide facilities such as HIRFL, HIAF, FAIR, FRIB, GANIL and RIKEN RIBF. Given the proper collision energy and centrality, the production yield of heavy FNDs is estimated to be $10^{-9}\sim10^{-4}$ per event (see SM for details), with equivalent cross sections $\sigma_{\mathrm{FND}}\approx 0.01~\mu\mathrm{b}\sim1~\mathrm{mb}$. Although the yields can provide sufficient statistics to search for FNDs, especially heavy ones in highly isospin asymmetric regions with low backgrounds, previous LGPT experiments likely missed FNDs due to detection blind spots: traditional $4\pi$ arrays typically resolve $Z$ rather than $A$ for heavy clusters, causing those with low $Z/A$ to be undistinguished from ordinary ones or rejected as double-hit background, while high-resolution fragment separators usually target forward-focused projectile fragmentation within narrow magnetic rigidity windows near the known drip line, missing mid-rapidity emissions with low $Z/A$ or rejecting extreme $B\rho$ outside standard acceptances. Combining thin targets with wide acceptances in fragment spectroscopy could overcome these limitations.

{\it Lifetimes.} 
In a standard first-order phase transition, true-vacuum bubbles nucleate and expand to convert the false vacuum on a timescale set by the nucleation and growth rates. In HICs, however, the ambient fireball expands so rapidly that the background medium enters the spinodal region and loses its potential barrier before bubbles can nucleate, fragmenting directly into nearly equal-sized true-vacuum fragments via spinodal decomposition \cite{borderie2001evidence,scavenius2001first}. Bubbles therefore do not nucleate in the background and thus do not engulf droplets. The bubble nucleation is hindered by the potential barrier inside droplets. The bubble nucleation rate $\Gamma$ away from the spinodal instability \cite{RN28} (see SM for details) and the resulting nucleation lifetime $\tau\sim1/(\Gamma V)$ for a droplet volume $V=\frac{4\pi}{3}R^3$ establish the wide $\tau \ge 1~\mathrm{\mu s}$ survival region, where FNDs outlive the $\mathcal{O}(\mathrm{ns})$ flight time to detectors. In the stable pocket, the enhanced finite-size effect in $\Omega_t$ generates a stable liquid state.

Other decay channels are likewise suppressed. First, the emission of light nuclides requires partial nucleation, which is suppressed by the same nucleation barrier above. Second, fission is mainly driven by the competition between the Coulomb repulsion and cohesive surface tension. Compared to the known fissile nuclides, the low charge-to-mass ratios drastically suppress the fissility parameter $x\propto Z^2/A$ of heavy FNDs, while the negative pressure further reinforces the suppression against deformation. Third, macroscopic fusion and nucleon absorption require secondary collisions, which are suppressed in HICs by the relativistic kinetic energy and the rapid spatial dilution. Fourth, unlike conventional dripping, where excess neutrons are nearly free to escape within $t\sim10^{-22}$ s, neutrons in an appreciable  part of heavy FNDs cannot freely penetrate the barrier with positive neutron separation energy $S_n$ as Fig. \ref{NZ} shows. They are confined by the negative pressure of the false vacuum, which can be understood by analogy with the quark confinement in the MIT bag model. Moreover, spontaneous neutron evaporation at finite $T$ is dynamically suppressed in HICs by the Pauli blocking under the local chemical equilibrium in neutron-rich environments.

In a neutron-rich environment, ambient neutrons can also bombard the droplet surface. These impinging neutrons need not enter the bulk structure; instead, they can act as inhomogeneous nucleation seeds that perturb the surface scalar field configuration and lower the effective potential barrier. Such surface seeds can catalyze the bubble nucleation \cite{blasi2026seeded,ignatius2001qcd} or trigger the surface evaporation and fragmentation, providing a plausible dominant decay channel in the presence of external impurities. If the production and seeded surface decay can be controlled, the catalyzed phase transition could potentially release controllable nuclear false vacuum energy.

{\it Discussion.} Incorporating gravity extends the FND to macroscopic astrophysical compact objects with equilibrium between the false vacuum and net electric charge. On the one hand, Coulomb repulsion counteracts the negative pressure of the false-vacuum nuclear matter. Whereas a conventional neutron star relies on a gravitationally bound crust across which the equation of state transitions smoothly to the vacuum, a stellar surface trapped in the false vacuum cannot maintain hydrostatic equilibrium with the external vacuum under negative nuclear pressure alone. By analogy with microscopic FNDs, a net charge contributing to the Einstein-Maxwell stress-energy tensor \cite{ray2003electrically} supplies an outward electrostatic pressure that balances this negative pressure, sustaining a false-vacuum surface at finite energy density to form a class of self-bound charged neutron stars.

On the other hand, the false vacuum binds net charges. Although in principle a neutron star can carry a net charge of $10^{20}$ C without significant changes in its structure \cite{arbanil2015equilibrium}, the practical upper bound is only $150$ C per solar mass \cite{krivoruchenko2018hydrostatic}, because gravity is much weaker than electrostatic repulsion that drives excess charged particles. In contrast, the negative pressure changes the boundary condition to $-P=\frac{1}{2}E^2=\frac{Q^2}{32\pi^2R^4}$, where $E$ is the electric field, $Q$ is the total net charge and $R$ is the stellar radius. Even a tiny negative pressure $-P\sim10^{-12}$ GeV$^4$ supports an electric field on the order of the Schwinger critical field $E\sim E_c\sim10^{18}$ V/m, corresponding to a net charge of $10^{16}$ C for $R\sim10$ km. Consequently, a net charge orders of magnitude above the conventional gravitational limit can be bound without disturbing the bulk stellar structure, sustaining surface electric fields up to $E \lesssim E_c$ that could drive electromagnetic signatures.

Finally, because false-vacuum trapping relies only on a first-order phase transition scenario, analogous droplets can emerge across other physical scales. Examples include quark nuggets \cite{Bai:2018dxf,Shao:2025tec} formed in first-order QCD phase transitions \cite{Shao2024short,Shaojd2024long} and metastable Fermi balls \cite{friedberg1977fermion,friedberg1977fermion2,xie2024revisiting} in cosmic phase transitions, which can serve as dark matter candidates or collapse into primordial black holes with distinct cosmological signatures \cite{hong2020fermi,lu2025black}.

{\it Acknowledgment.} We thank Y. G. Ma, F. S. Zhang and Y. P. Zhang for helpful discussions. This work was supported in part by the National Natural Science Foundation of China (NSFC) Grant Nos. 12235016 and 12221005.

\bibliographystyle{unsrt}
\bibliography{main}
\end{document}


\title{Supplemental Material}
\author{Jingdong Shao}
\email[]{shaojingdong19@mails.ucas.ac.cn} 
\affiliation{School of Physical Sciences, University of Chinese Academy of Sciences, Beijing 101408, China} 
\author{Mei Huang}
\email[]{huangmei@ucas.ac.cn (corresponding author)}
\affiliation{School of Nuclear Science and Technology, University of Chinese Academy of Sciences, Beijing 101408, China}	
\begin{abstract}
     
\end{abstract}
\maketitle
In this Supplemental Material, we provide the detailed formulation of the two-flavor nonlinear Walecka model and the three-dimensional liquid-gas phase transition (LGPT) phase diagram. We also present the parameterization of the isospin- and order-parameter-dependent surface tension and the formulation of the neutron separation energy. Next we show radii of false-vacuum-trapped neutron-rich droplets (FNDs) and estimate the production yields. We finally present the evaluations of the Euclidean bounce action of the bubble nucleation rate $\Gamma$ at finite temperature $T\sim5$ MeV and at $T=0$.

\section{Walecka Model}
The Lagrangian of the two-flavor relativistic mean-field Walecka model describing nucleons $\Psi_N$ ($N=p,n$) coupled to the scalar meson $\sigma$ and vector meson $\omega^\mu$ reads \cite{WALECKA1974491}
\begin{equation}
\mathcal{L}=\sum_N\bar\Psi_N(i\partial^\mu\gamma_\mu-m_N+g_\sigma\sigma-g_\omega\omega^\mu\gamma_\mu)\Psi_N+\frac{1}{2}(\partial \sigma)^2+\frac{1}{2}m^2_\omega\omega^\mu\omega_\mu-\frac{1}{4}(\partial_{\mu}\omega_{\nu}-\partial_{\nu}\omega_{\mu})(\partial^{\mu}\omega^{\nu}-\partial^{\nu}\omega^{\mu})-U(\sigma),
\end{equation}
with the nonlinear scalar self-interaction potential
\begin{equation}
  U(\sigma)=\frac{1}{2}m^2_\sigma\sigma^2+\frac{1}{3}bm_N(g_\sigma\sigma)^3+\frac{1}{4}c(g_\sigma\sigma)^4.
\end{equation}
We set the nucleon and meson masses to $m_N=939$~MeV, $m_\sigma=550$~MeV, and $m_\omega=783$~MeV, and fix the coupling constants to $g_\sigma=8.96$, $g_\omega=9.24$, $b=0.00692$, and $c=-0.0048$ following Ref.~\cite{he2023speed} by fitting the compression modulus, symmetry energy, effective nucleon mass, and binding energy at nuclear saturation density. In the mean-field approximation, the grand potential density is given by
\begin{equation}
\begin{aligned}
  \Omega &= \frac{1}{2}m^2_\sigma\sigma^2-\frac{1}{2}m^2_\omega\omega^\mu\omega_\mu+\frac{1}{3}bm_N(g_\sigma\sigma)^3+\frac{1}{4}c(g_\sigma\sigma)^4 \\
  &\quad -2T\sum_{i=p,n}\int\frac{d^3\vec p}{(2\pi)^3}\ln\left[\left(1+e^{-\frac{E_i+\mu_i^*}{T}}\right)\left(1+e^{-\frac{E_i-\mu_i^*}{T}}\right)\right],
\end{aligned}
\end{equation}
where $\vec p$ is the nucleon momentum and $E_i=\sqrt{m^2+\vec p^2}$ is the single-particle energy with the effective nucleon mass $m=m_N-g_\sigma\sigma$. The index $i$ labels protons ($p$) and neutrons ($n$), and the effective chemical potential reads $\mu_i^*=\mu_i-g_\omega\omega_0$, where $\mu_n=\mu+\mu_I$ and $\mu_p=\mu-\mu_I$ incorporate the baryon chemical potential $\mu$ and isospin chemical potential $\mu_I$. The effective nucleon mass $m$ is substantially reduced in the liquid phase and restored toward $m_N$ in the dilute gas phase. Specifically, the grand potential of the free neutron gas with the restored mass $m_N$ reads

\begin{equation}
   \Omega_n=-2T\int\frac{\mathrm{d}^3\vec p}{(2\pi)^3}\left(\ln\left(1+e^{-\frac{\sqrt{m_N^2+p^2}+\mu_n}{T}}\right)+\ln\left(1+e^{-\frac{\sqrt{m_N^2+p^2}-\mu_n}{T}}\right)\right).
\end{equation}

\begin{figure}
    \centering
    \includegraphics[width=0.5\linewidth]{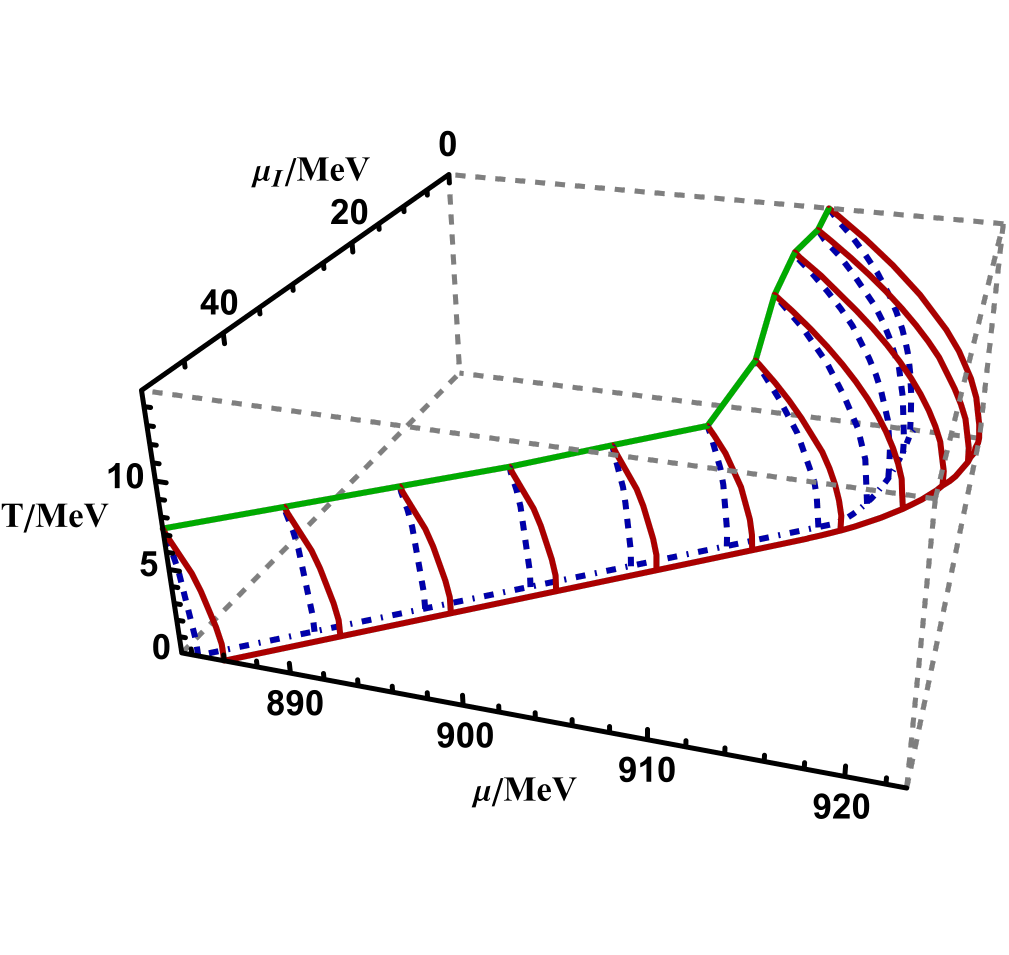}
\caption{Three-dimensional LGPT phase diagram in the $(T, \mu, \mu_I)$ space. The red solid curves denote the static first-order phase transition lines across different slices, while the blue dashed curves denote the corresponding liquid spinodal lines. The green curve traces the trajectory of the critical endpoints.}
    \label{phase3D}
\end{figure}
\begin{figure}
    \centering
\includegraphics[width=0.5\linewidth]{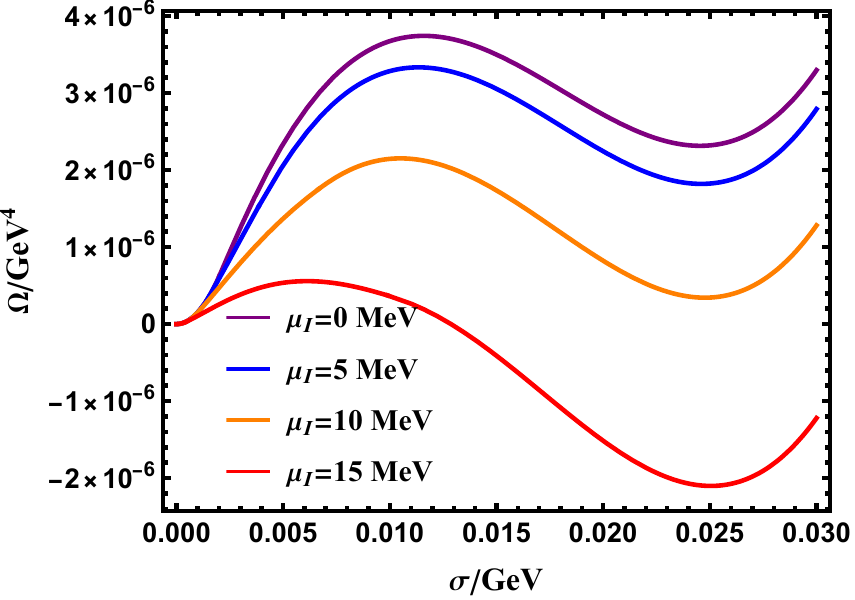}
   \caption{Potential curves at $T=0$, $\mu=921$ MeV and different $\mu_I$.}
    \label{potential_mui}
\end{figure}

The phase diagram is determined by solving the coupled gap equations $\partial\Omega/\partial\sigma = 0$ and $\partial\Omega/\partial\omega_0 = 0$. Figure \ref{phase3D} shows the three-dimensional phase diagram of the LGPT in the $(T,\mu,\mu_I)$ space. The red solid curves denote the static first-order phase transition lines across different slices, while the blue dashed curves denote the spinodal lines. The green curve traces the trajectory of the critical endpoints. As the isospin chemical potential $\mu_I$ increases from the symmetric limit $\mu_I = 0$ to the neutron-rich regions, the isospin asymmetry shifts both the first-order phase transition lines and liquid spinodal  lines toward lower baryon chemical potential $\mu$. The three-dimensional wedge-shaped space enclosed by the red and blue surfaces indicates the extended metastable false vacuum window of isospin-asymmetric nuclear matter, where the isospin asymmetry shifts the grand potential as Fig. \ref{potential_mui} shows. The potential curves at $T=0$, $\mu=921$ MeV and different $\mu_I$ are plotted in Fig. \ref{potential_mui} as examples. As $\mu_I$ increases, the phase transition is effectively retarded at fixed $\mu$ by the isospin asymmetry in the neutron-rich region.

\section{Surface tension}
Following Lattimer and Swesty \cite{1991NuPhA}, we parameterize the isospin-dependent surface tension as
\begin{equation}
  \sigma_S=\sigma_0\frac{2^{p+1}+b_S}{b_S+Y_P^{-p}+(1-Y_P)^{-p}},
\end{equation}
where $Y_P=Z/A=Z/(Z+N)$ denotes the proton fraction of a droplet with $A$ baryons ($Z$ protons and $N$ neutrons). $\sigma_0$ denotes the surface tension of symmetric infinite nuclear matter and $b_S$ determines the isospin dependence. The parameters quantitatively vary among models, but the qualitative behavior is model-independent: the surface tension decreases with increasing isospin asymmetry and drops to zero for pure neutrons \cite{centelles1998semiclassical,danielewicz2003surface}. The surface tension remains within the same order of magnitude across different models \cite{steiner2005isospin}, and we adopt the baseline parameters $b_S=13.33$, $p=3$ and $\sigma_0=4.28\times10^{-5}$ GeV$^3$ following \cite{newton2013survey}.

Although $\sigma_S$ is parameterized for semi-infinite nuclear matter in the liquid phase, it provides a reliable approximation for finite droplets under the thin-wall approximation. However, evaluating the total grand potential density $\Omega_t(\sigma)$ across intermediate field configurations during the phase transition requires the explicit form of the order-parameter-dependent surface tension. Under the thin-wall approximation, the intermediate field configuration $\sigma(r)$ between the interior value $\sigma$ and the exterior one $\sigma = 0$ takes the kink form 
\begin{equation}
  \sigma(r) = \frac{\sigma}{2}\left[1 - \tanh\left(\frac{r-R}{D}\right)\right],
\end{equation}
where $D$ is the wall thickness. The gradient contribution to the effective surface tension $\sigma_{\mathrm{eff}}(\sigma)$ is
\begin{equation}
  \sigma_{\mathrm{eff}}(\sigma) = \int \mathrm{d}r\,\frac{K}{2}\left(\frac{\mathrm{d}\sigma(r)}{\mathrm{d}r}\right)^2 = \frac{K\sigma^2}{6D} \simeq \sigma_S\left(\frac{\sigma}{\sigma_l}\right)^2,
\end{equation}
where $K$ is the effective coefficient and $\sigma_l$ is the expectation value of the order parameter of $\Omega$ in infinite nuclear matter. Because the effective surface tension should match the value in the semi-infinite nuclear matter $\sigma_{\mathrm{eff}}(\sigma_l) = \sigma_S$, the unknown parameters are absorbed into the known surface tension $K/(6D) \simeq \sigma_S/\sigma_l^2$, which gives the leading-order quadratic term $(\sigma/\sigma_l)^2$.

\section{Neutron Separation Energy}

Within the thermodynamic approximation, the total energy $E_A(Z,N)$ of one FND consists of the bulk thermodynamic energy, the surface energy, and the Coulomb energy,
\begin{equation}
  E_A(Z,N) = -\frac{4\pi}{3}R^3 P_{\mathrm{in}} + \mu_n N + \mu_p Z + 4\pi R^2\sigma_{\mathrm{eff}} + \frac{3Z^2 e^2}{20\pi R},
\end{equation}
evaluated at the mechanical equilibrium radius $R$. Fixing the proton number $Z$, the single-neutron separation energy required to emit one neutron into the exterior true vacuum is defined as
\begin{equation}
  S_n(Z,N) = E_A(Z,N-1) + m_N - E_A(Z,N).
\end{equation}
A positive separation energy $S_n > 0$ indicates that spontaneous single-neutron emission is energetically forbidden, whereas $S_n < 0$ indicates thermodynamic instability toward neutron dripping. Because the mechanical equilibrium condition eliminates the leading $\partial E_A/\partial R$ variation, the energy difference for heavy FNDs ($A\gg1$) reduces to $E_A(Z,N) - E_A(Z,N-1) \approx (\partial E_A/\partial N)_Z \approx \mu_n $, where the second approximate equality neglects the isospin dependence of $\sigma_{\mathrm{eff}}$. Consequently, $S_n \approx m_N - \mu_n$ for heavy FNDs closely matches the direct thermodynamic criterion $m_N - \mu_n > 0$. For light FNDs containing only a small number of nucleons, however, the continuous thermodynamic approximation neglects discrete single-particle levels, shell and pairing effects, and the calculated $S_n$ in the low-mass region should be regarded as a qualitative extrapolation rather than a quantitative prediction.

\section{Radii and Estimate of FND Production Yields}
\begin{figure}
  \centering
\includegraphics[width=0.6\linewidth]{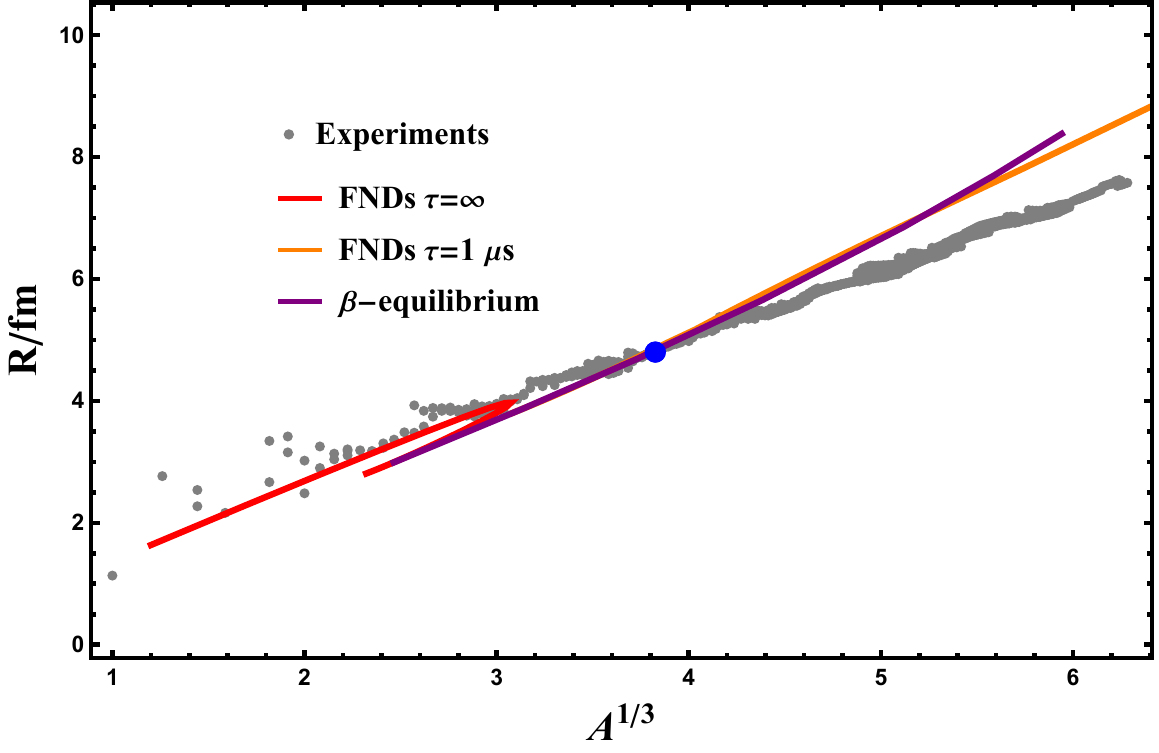}
  \caption{Radii of droplets compared to the equivalent sharp radii $R\approx\sqrt{\frac{5}{3}}r_{\mathrm{ch}}$ inferred from experimental charge radii $r_{\mathrm{ch}}$.}
  \label{Rad}
\end{figure}
Figure \ref{Rad} compares droplet radii $R$ to the equivalent sharp radii $R\approx\sqrt{\frac{5}{3}}r_{\mathrm{ch}}$ inferred from experimental charge radii $r_{\mathrm{ch}}$ \cite{Angeli:2013epw}. The orange curve marks FNDs with lifetimes $\tau=1~\mathrm{\mu s}$, while the red one marks those with lifetimes $\tau=\infty$. The purple curve marks FNDs with $\beta$-equilibrium. The relation $r_{\mathrm{ch}}\approx\sqrt{\frac{3}{5}}R$ follows from approximating the droplets as uniformly charged spheres. Generally, the droplet radii are similar to those of known nuclides with the same baryon numbers and thus suggest similar sub-saturation baryon number density. The blue point marks the intersection point of the droplet radii and experimental data around $A\approx56$. However, for heavy droplets, e.g., $A^{1/3}>4.5$, the droplet radii exceed those of known nuclides by around $1$ fm. The possible larger radii may lead to larger cross sections of secondary collisions and thus stronger hydrodynamic drag in the surrounding medium. The resulting anomalously short mean free paths could provide evidence in the search for droplets, analogous to the searches for dense anomalous matter via anomalously long mean free paths \cite{friedlander1980evidence,friedlander1983anomalous}. 

The production yield of heavy FNDs per event in $\mathrm{Au}+\mathrm{Au}$ collisions at intermediate energies $E_{\mathrm{lab}}\sim$ a few hundred AMeV can be estimated by
\begin{equation}
Y_{\mathrm{FND}} \simeq Y_{\mathrm{dyn}}(A) \times P_{\mathrm{iso}}(\delta) \times P_{\mathrm{ex}}.
\end{equation}

The first factor $Y_{\mathrm{dyn}}$ is the production yield of early dynamical emission per event and is estimated based on that of the heavy intermediate-mass fragments (IMFs, $10 < Z \le 20$). Geometrically averaging the ALADIN multiplicity curve over the reduced impact parameter gives a total IMF ($20\ge Z\ge3$) yield \cite{Schuttauf1996} 
\begin{equation}
    Y_{\mathrm{tot}}\approx1.8\sim2.5 \,\,\mathrm{per\,\,event},
\end{equation}
and the empirical power-law charge spectrum $Y(Z) \propto Z^{-\zeta}$ (with the index $\zeta \approx 2.0\sim2.4$) \cite{Kreutz1993} isolates the heavy IMF ($10 < Z \le 20$) component 
\begin{equation}
    Y_{\mathrm{heavy}} = Y_{\mathrm{tot}} \left(\sum_{Z=11}^{20} Z^{-\zeta} / \sum_{Z=3}^{20} Z^{-\zeta}\right) \approx 0.06\sim0.16\,\,\mathrm{per\,\,event}.
\end{equation}
The early dynamical decoupling fraction $f_{\mathrm{dyn}} \approx 25\%\sim40\%$ obtained in quantum molecular dynamics simulations coupled with simulated annealing clusterization \cite{Gossiaux1997} gives
\begin{equation}
    Y_{\mathrm{dyn}} = f_{\mathrm{dyn}}Y_{\mathrm{heavy}} \approx 0.02\sim0.06\,\,\mathrm{per\,\,event},
\end{equation}
which corresponds to a probability of 
\begin{equation}
    P_{\mathrm{dyn}} = 1 - \exp(-Y_{\mathrm{dyn}}) \approx 2\%\sim6\%
\end{equation}
via Poisson statistics.

The second factor $P_{\mathrm{iso}}$ is the probability of reaching an isospin asymmetry. The asymmetry $\delta \equiv (N-Z)/A$ at freeze-out is modeled by a Gaussian distribution \cite{PhysRevC.80.054606,PhysRevC.75.044605}
\begin{equation}
P(\delta|A) = \frac{1}{\sqrt{2\pi}\sigma_\delta} \exp\!\left[-\frac{(\delta - \bar{\delta})^2}{2\sigma_\delta^2}\right], \quad \sigma_\delta^2 = \frac{T}{2 A E_{\mathrm{sym}}},
\end{equation}
where the variance $\sigma_\delta$ is governed by the ratio of the effective temperature $T\approx5$ MeV \cite{PhysRevLett.75.1040} to the symmetry energy $E_{\mathrm{sym}}\approx15$ MeV at sub-saturation density \cite{PhysRevC.76.024606}. The local mean of $\delta$ is $\bar{\delta} \approx 0.2$ for $\mathrm{Au}+\mathrm{Au}$, and integrating the Gaussian tail
\begin{equation}
    P(\delta > \delta_c | A) = \frac{1}{2}\,\mathrm{erfc}[(\delta_c - \bar{\delta})/(\sqrt{2}\sigma_\delta)]
\end{equation}
gives the conditional probabilities of reaching a certain isospin asymmetry. For the IMFs with $A\sim40$, the probabilities are $P_{\mathrm{iso}}\approx0.06,\,10^{-3}\mathrm{\,and\,}10^{-6}$ for $\delta_c=0.3,\,0.4\,\mathrm{\,and\,}0.5$ respectively.

The third factor $P_{\mathrm{ex}}$ is the probability of exciting one cluster across the potential barrier into the false vacuum. The excitation occurs in an early stage $T\sim\mathcal{O}(10)$ MeV and the excitation energy $\Delta E_{\mathrm{ex}}$ can be taken as the energy scale of the LGPT $\Delta E_{\mathrm{ex}}\sim\mathrm{a\,\,few}\times10$ MeV, thus an excitation factor gives $P_{\mathrm{ex}} \sim \exp(-\Delta E_{\mathrm{ex}} / T) \sim 0.1$. 

Combining these factors yields $Y_{\mathrm{FND}}\approx10^{-9}\sim10^{-4}$ per event, depending on the isospin asymmetry and sizes. This production yield corresponds to a probability $P_{\mathrm{dyn}}P_{\mathrm{iso}}P_{\mathrm{ex}}\approx10^{-9}\sim10^{-4}$ of generating one heavy FND in one event, or to a production cross section $\sigma_{\mathrm{FND}} \approx Y_{\mathrm{FND}}\sigma_{\mathbf{Au}}\approx 0.01~\mu\mathrm{b}\sim1~\mathrm{mb}$, where $\sigma_{\mathbf{Au}}\sim\mathcal{O}(10)$ b is the cross section of $\mathrm{Au}+\mathrm{Au}$ collisions. The production yield and cross section are then accessible to experiments in HIRFL and HIAF, however, they are estimated by known data of IMFs and geometrical average, and the proper collision conditions including the collision energy and centrality require future searches.

\section{Bounce Action and Droplet Lifetimes}
During a first-order phase transition, bubbles nucleate, expand and convert the false vacuum on a timescale governed by the nucleation and expansion rates. The nucleation rate is determined by the Euclidean action of a bounce solution of the field configuration inside and outside bubbles. The false vacuum corresponds to the liquid phase with the total potential $\Omega_t$, while the true vacuum corresponds to the dilute gas phase with $\Omega_t=0$.

At zero temperature, quantum tunneling dominates in the absence of thermal fluctuations, and bubbles therefore form through quantum nucleation. The configuration of field $\sigma(r)$ is given by the bounce solution to an $O(4)$-symmetric equation of motion
\begin{equation}
    \frac{\mathrm{d}^2\sigma}{\mathrm{d}r^2}+\frac{3}{r}\frac{\mathrm{d}\sigma}{\mathrm{d}r}=\frac{\mathrm{d}\Omega_t}{\mathrm{d}\sigma}
\end{equation}
with boundary conditions $\left.\frac{\mathrm{d}\sigma}{\mathrm{d}r}\right|_{r=0}=0$ and $\sigma(\infty)\rightarrow$false vacuum. The four-dimensional Euclidean bounce action $S_4$ is then evaluated on this solution
\begin{equation}
    S_4=2\pi^2\int \mathrm{d}rr^3(\frac{1}{2}(\nabla \sigma(r))^2+\Omega_t(\sigma(r))).
\end{equation}
At finite temperature, the thermal fluctuations prevail, and thus bubbles form through thermal nucleation. The symmetry in the time dimension is violated by temperature and the $O(4)$-symmetric bubble reduces to an $O(3)$-symmetric one 
\begin{equation}
    \frac{\mathrm{d}^2\sigma}{\mathrm{d}r^2}+\frac{2}{r}\frac{\mathrm{d}\sigma}{\mathrm{d}r}=\frac{\mathrm{d}\Omega_t}{\mathrm{d}\sigma}
\end{equation}
with boundary conditions $\left.\frac{\mathrm{d}\sigma}{\mathrm{d}r}\right|_{r=0}=0$ and $\sigma(\infty)\rightarrow$false vacuum. The bounce action reduces to the three-dimensional $\frac{S_3}{T}$, where
\begin{equation}
    S_3=\int \mathrm{d}^3r(\frac{1}{2}(\nabla \sigma(r))^2+\Omega_t(\sigma(r))).
\end{equation}

The nucleation rate away from spinodal instability is \cite{RN28}
\begin{equation}
\Gamma=\left\{\begin{array}{lr}
  T^4\left(\frac{S_3}{2\pi T}\right)^\frac{3}{2}e^{-S_3/T},\,\, \mathrm{if\,\,} T>0 \\
  \\
   B^4\left(\frac{S_4}{2\pi}\right)^2e^{-S_4}, \,\,\mathrm{if\,\,} T\sim0 \end{array}
\right.
\label{nucleation_rate}
\end{equation}
where $B$ is the dimensional factor that denotes the scale of interaction and could be taken as the inverse scale of the system $\sim R^{-1}$. The lifetime for a certain volume $V$ is estimated to be the time for one bubble to nucleate $\tau=\frac{1}{\Gamma V}$.

Conservatively, Eq.~(\ref{nucleation_rate}) is valid for $S_4>10$ or $S_3/T>10$, but breaks down near the spinodal instability regime, where $S_4\sim\mathcal{O}(1)$ or $S_3/T\sim\mathcal{O}(1)$. The transition rate is governed by the dynamical relaxation time $\tau_{\mathrm{dyn}}$ rather than the potential barrier, and the phase conversion proceeds via spinodal decomposition. In HICs, the fireball expansion time is of order $\mathcal{O}(100)$~fm/$c$, which requires the nucleation rate $\Gamma\sim 10^{-4}$~fm$^{-4}$  for true vacuum bubbles to nucleate within this timescale. For $B\sim 10$~fm$^{-1}$, the required bounce action $S_4\sim 0.031$ essentially corresponds to the spinodal instability, which confirms the spinodal decomposition scenario during the fireball expansion.

\begin{figure}
    \centering
\includegraphics[width=0.5\linewidth]{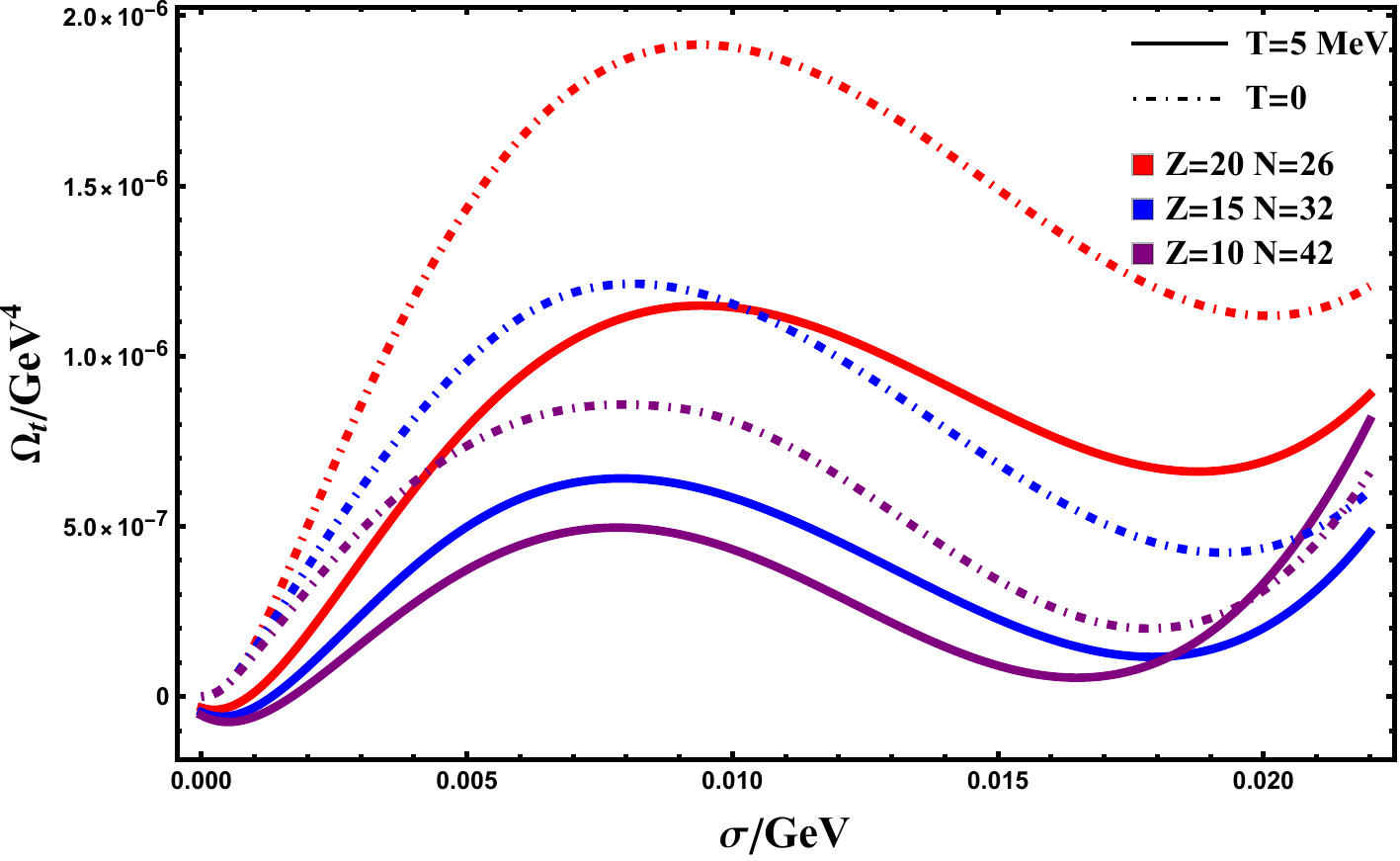}
    \caption{Potential curves of different droplets at $T=0$ (dashed) and $T=5$ MeV (solid).}
    \label{potential}
\end{figure}

\begin{figure}
    \centering
    \includegraphics[width=0.5\linewidth]{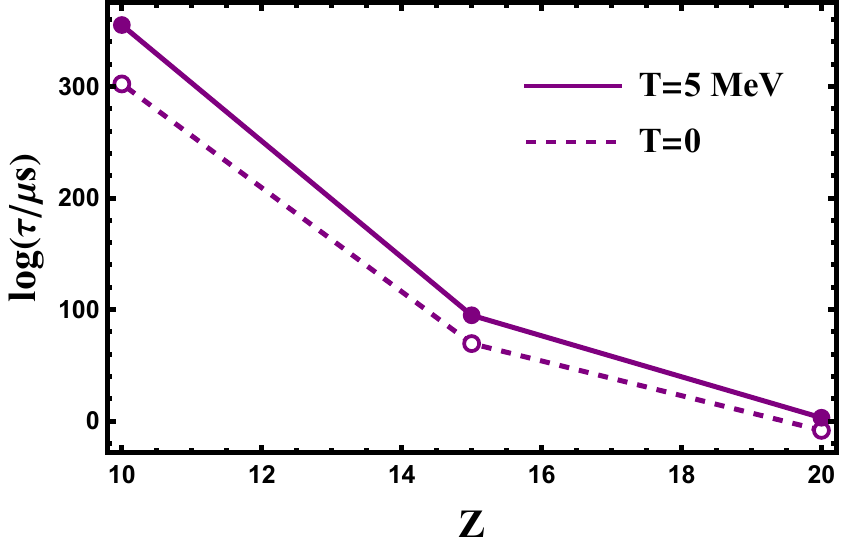}
    \caption{Lifetimes of different droplets at $T=0$ (dashed) and $T=5$ MeV (solid).}
    \label{life}
\end{figure}

Figure \ref{potential} shows the potential curves of three droplets with different numbers of protons and neutrons at $T=0$ (dashed) and $T=5$ MeV (solid) as examples. At finite temperature, the difference of the grand potential between the true and false vacuum $\Delta\Omega_t$ is insufficient to overcome the potential barrier at a short time; as temperature decreases, $\Delta\Omega_t$ increases but the potential barrier also rises, and the net effect is shown in Fig. \ref{life} by comparing the corresponding lifetimes of these three droplets at $T=0$ and at $T=5$ MeV. Lifetimes at $T=5$ MeV are larger than those at $T=0$, which is consistent with the fact that the region of droplets with $\tau\ge1~\mathrm{\mu s}$ at $T=5$ MeV contains the one at $T=0$. We expect that droplets with lifetimes $\tau\ge1$ $\mathrm{\mu s}$  are able to survive to detectors and thus temperature does not compromise the region with detectable lifetimes.
\bibliography{main}